\documentclass[]{article}
\usepackage{emulateapj,onecolfloat}

\def\eg{{\it e.g.}}
\def\etal{{\it et al.}}
\def\etc{{\it etc.}}
\def\ie{{\it i.e.}}
\def\Lres{L_{\rm res}}
\slugcomment{\it Revised version submitted to ApJ}

\begin{document}

\twocolumn[
\title{Bar-Halo Friction in Galaxies II: Metastability}
\author{J. A. Sellwood}
\affil{Rutgers University, Department of Physics \& Astronomy, \\
       136 Frelinghuysen Road, Piscataway, NJ 08854-8019 \\
       {\it sellwood@physics.rutgers.edu}}
\and

\author{Victor P. Debattista\altaffilmark{1}}

\affil{Astronomy Department, University of Washington, \\
Box 351580, Seattle, WA 98195-1580 \\
{\it debattis@astro.washington.edu}}
\altaffiltext{1}{Brooks Fellow}

\begin{abstract}
It is well-established that strong bars rotating in dense halos
generally slow down as they lose angular momentum to the halo through
dynamical friction.  Angular momentum exchanges between the bar and
halo particles take place at resonances.  While some particles gain
and others lose, friction arises when there is an excess of gainers
over losers.  This imbalance results from the generally decreasing
numbers of particles with increasing angular momentum, and friction
can therefore be avoided if there is no gradient in the density of
particles across the major resonances.  Here we show that anomalously
weak friction can occur for this reason if the pattern speed of the
bar fluctuates upwards.  After such an event, the density of resonant
halo particles has a local inflexion created by the earlier exchanges,
and bar slowdown can be delayed for a long period; we describe this as
a metastable state.  We show that this
behavior in purely collisionless $N$-body simulations is far more
likely to occur in methods with adaptive resolution.  We also show
that the phenomenon could arise in nature, since bar-driven gas inflow
could easily raise the bar pattern speed enough to reach the metastable
state.  Finally, we demonstrate that mild external, or
internal, perturbations quickly restore the usual frictional drag, and
it is unlikely therefore that a strong bar in a galaxy having a dense
halo could rotate for a long period without friction.
\end{abstract}

]\addtocounter{footnote}{-1}

\section{Introduction}
\label{intro}
In earlier work (Debattista\footnote{Brooks Fellow} \& Sellwood 1998, 2000, hereafter DS98 \&
DS00), we showed that strong, rapidly rotating bars in dense halos
slow down quickly due to dynamical friction.  Our finding of strong
friction is consistent with theoretical work (\eg\ Weinberg 1985;
Hernquist \& Weinberg 1992, Weinberg 2004) and the consequent braking
of bars is reported in other fully self-consistent simulations (\eg\
Little \& Carlberg 1991; O'Neill \& Dubinski 2003).

Most strong bars in real galaxies appear to rotate rapidly, in the
sense that the dimensionless ratio ${\cal R} \la 1.4$; here ${\cal R}
\equiv R_c/a_B$, with $R_c$ being the radius of corotation and $a_B$
being the semi-major axis of the bar.\footnote{It is the value of this
ratio today that matters, and not the fact that the bar has slowed a
lot, as was mis-stated by Athanassoula (2003).  Finding evidence of
the past history of the bar pattern speed would be an even greater
observational challenge.\par} This ratio is not easy to determine
directly from observation, but Aguerri \etal\ (2003) summarize results
from a number of galaxies and Debattista \& Williams (2004) add a new
result with much lower uncertainty obtained by an integral-field
method (but see also Rautiainen, Salo \& Laurikainen 2005).  Indirect
evidence comes from the location of dust lanes (Athanassoula 1992) and
rings (Buta \& Combes 1996).

In DS98 and DS00, we combined our finding that bars in dense halos
soon become slow with the observation that bars in real galaxies are
fast to attempt to constrain the contribution to the central
attraction that comes from the inner dark matter halo.  But others
have challenged the result that bars become slow, which calls the
constraint on halo density into question.  The criticism by
Athanassoula (2003) is more of interpretation than substance, since
strong bars are subject to fierce braking in her simulations, and bars
that experience little friction are weak.  We will report $\cal R$
values for models similar to hers in paper III of this series
(Sellwood \& Debattista 2006).

Valenzuela \& Klypin (2003, hereafter VK03), on the other hand, claim
counter-examples of strong bars in dense halos that stay fast for
cosmologically interesting periods of time.  VK03 argue that their
different result, which disagrees with all previous simulations and
with the above-cited theoretical work, is in fact correct, and suggest
that only their simulations have the numerical resolution to reveal
the proper behavior.

When started from their initial conditions, simulations with our
code (\S\ref{results}) behave in many respects as reported by VK03;
they form strong bars, of similar lengths and pattern speeds, for
example.  However, the bars in our simulations generally exhibit
strong friction and quickly become unacceptably slow.  One of our
experiments is anomalous, however, and shows an even longer delay in
the onset of friction than found by VK03.  Although our anomalous
result may be artificial (it is not reproducible when numerical
parameters are changed), it presented us with an opportunity to
discover how friction can be avoided for long periods.

Tremaine \& Weinberg (1984) showed that dynamical friction in a
quasi-spherical system arises because of resonant interactions between
a rotating potential perturbation and the orbits of particles.  As the
decreasing bar pattern speed sweeps across a resonance with some halo
orbits, their angular momenta may be substantially changed (Sellwood
2005, hereafter Paper I).  Halo particles may either gain or lose
angular momentum as they cross a resonance, and to first order there
is no net loss or gain.  However, to second order in the perturbing
potential, there is usually a net gain in angular momentum by the halo
particles, leading to a friction-like drag on the bar even in a
perfectly collisionless system.  The bias
arises because the number density of halo particles is usually a
decreasing function of angular momentum, leading to the excess of
gainers over losers.

The distribution of particles about the principal resonances
responsible for the torque on the bar is altered by the evolution
itself, as shown in Paper I.  In particular, exchanges between the
halo particles and the bar combine with the time-dependent pattern
speed, to cause the density of particles in the vicinity of the
resonance to develop a pronounced shoulder, with the instantaneous
resonance lying on the high angular momentum side of the shoulder
where the gradient is negative.

In \S\ref{metastability}, we show that should $\Omega_p$ fluctuate
upwards, for some reason, after a period of normal friction, the
resonance may cross to the other side of the previously-created
shoulder where the gradient with angular momentum may be locally flat,
or even reversed.  In these circumstances, net exchanges at the
resonance will no longer lead to friction.  Since friction is
generally dominated by a few resonances, which all behave in a similar
manner, such a change can lead to a dramatic decrease in the net
torque.\footnote{Holley-Bockelmann \& Weinberg (2005) construct a halo
with no gradient at the outset, and report mild friction on the
bar, but such models are quite contrived.}

We show that an upward fluctuation in the bar pattern speed in
collisionless $N$-body simulations is not reproducible when numerical
parameters are changed.  The pattern speed fluctuates upward before
periods of weak friction in experiments A$_1$ and B reported by VK03.
In \S\ref{artifact} we show that adaptive mesh refinement is the
likely culprit for a numerical artifact causing their anomalous
results.

The possibility that bars in real galaxies with dense halos could
experience little friction would be of great interest if there were a
physically realistic reason for the bar pattern speed to rise.  We
show in \S\ref{gasinflow} that gas inflow in bars may have such a
consequence, but we also find (\S\ref{fragile}) that other physically
relevant factors are likely to prevent the low-friction state from
persisting for long.

We describe the low-friction state as metastable both because it
relies on a local feature in the phase-space density and because it is
fragile.  It should be noted that metastability applies only to
anomalously weak friction on a strong bar in a dense halo -- friction
will always be mild when the bar is weak or the halo density low.

\begin{figure}[t]
\plotone{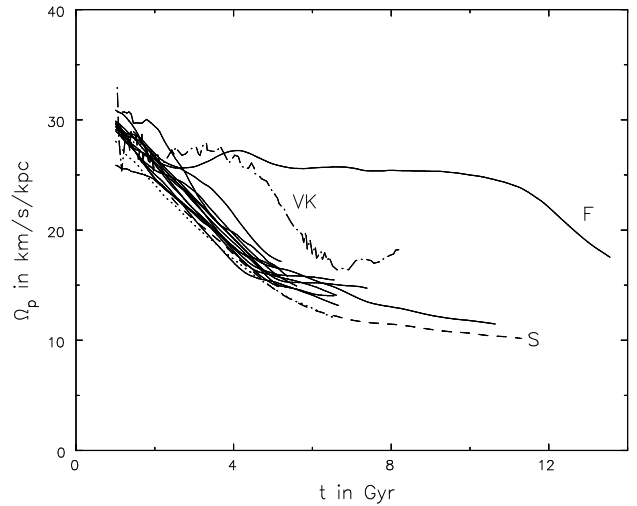}
\caption{\footnotesize The pattern speed evolution in many experiments
using our code that were all begun from the initial model A$_1$ of VK03.
Each line shows a separate simulation that differs from the others
only in the numerical parameters adopted.  Curves from our runs F and
S are so labeled, and the dot-dash line labeled VK shows the
evolution, reproduced from Fig.~10 of VK03, when this model was
simulated with their ART code.}
\label{omevol}
\end{figure}
\begin{table}[b]
\caption{Numerical parameters for Runs F \& S}
\label{params}
\begin{tabular}{@{}lrr}
                   & Cylindrical grid       & Spherical grid \\
\hline
Grid size          & $(N_R,N_\phi,N_z)\quad$ \\
                   &  $ = (81,128,125)$     & $N_r = 300$ \\
Angular compnts    & $0\leq m \leq 8$       & $0 \leq l \leq 4$ \\
Outer radius       & 29.4 kpc               & 350 kpc \\
$z$-spacing        & 21.875 pc \\
Softening length $\epsilon$ & 43.75 pc  \\
Number of particles & 199\,996               & 3\,351\,798 \\
Shortest time step & 0.14 Myr               & 0.14 Myr \\
\hline
\end{tabular}
\end{table}

\vfill\eject
\section{Simulations}
\label{results}
Model A$_1$ presented by VK03 had an exponential disk embedded in a
cosmo\-logically-motivated, cold dark matter halo that had an
approximately NFW density profile (Navarro, Frenk \& White 1997) with
a concentration index $c\approx15$, after compression.  These authors
kindly made available the initial positions, velocities, and masses of
all the particles in their model and we have run a large number of
simulations from these initial conditions using our hybrid, polar-grid
code (Sellwood 2003).  The numerical parameters used for two runs
(runs F \& S explained below) are given in Table~1, and we varied the
numerical parameters in other runs.  We employ the units used by VK03;
the exponential disk has a mass of $4.28 \times 10^{10}\;$M$_\odot$
and scale length of 3.5~kpc; the halo has a mass of $2.00 \times
10^{12}\;$M$_\odot$.

The evolution of the pattern speed, $\Omega_p$, in many of our
simulations using their initial particles, is shown in
Figure~\ref{omevol}; the dot-dashed curve shows the result reported by
VK03 for this model using their ART code (Kravtsov, Klypin \& Khokhlov
1997).  In contrast to their result, the bars in all but one of our
cases slow quite quickly, reaching $\sim 16\;$km/s/kpc by $\sim
5\;$Gyr, whereas VK03 found that the pattern speed remained roughly
constant, $\Omega_p \simeq 27\;$km/s/kpc, until $\sim4\;$Gyr and then
decreased below 20~km/s/kpc by $\sim 6.5\;$Gyr.  In fact, the
slow-down {\it rate} in all experiments is remarkably similar, the
different curves being approximately parallel, with a generally small
time offset.  Apart from the delay, the decrease reported by VK03 is
roughly consistent with the drop we observe in most cases.

However, in one of our experiments, which we denote run F (for fast),
$\Omega_p$ did not decrease significantly until almost 12~Gyr from the
start, although when it did start to decrease, the rate of decline was
again quite comparable to that in the majority of our experiments, and
that found by VK03.  As already noted, our experiments differ from
each other only by the numerical parameters adopted.  The parameters
for run F (Table 1) are typical of all the experiments shown and the
different behavior is not a question of a lack of numerical
convergence; parameters (time step, grid spacing, softening, \etc) in
the other experiments were both refined, and made coarser, from the
set that led to the anomalous result.

In all but one case, we used all the particles supplied by VK03; in
the remaining case (shown by the dotted line), we employed the full
number of disk particles but only every tenth halo particle, which we
made ten times more massive.  The result, even with this quite drastic
reduction in numerical quality, is that the bar slowed in a similar
manner as shown.

\begin{figure}[t]
\plotone{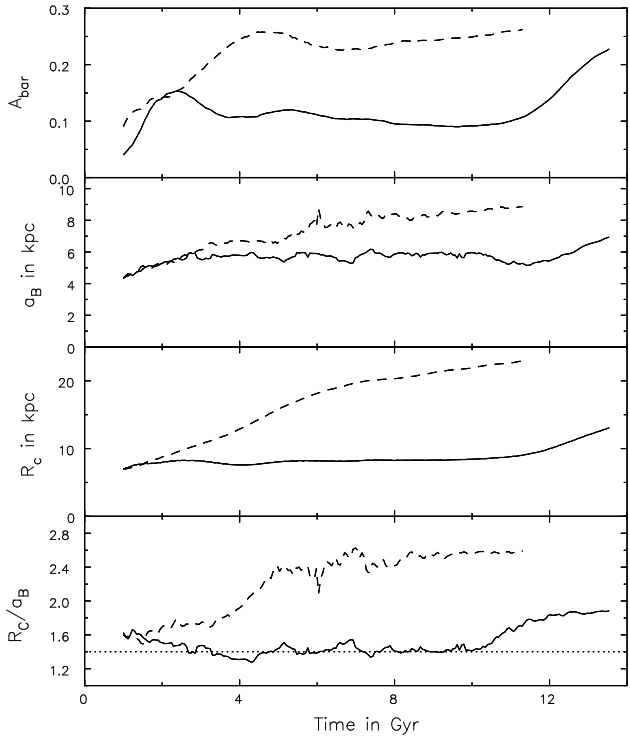}
\caption{\footnotesize From top to bottom, the bar amplitude, the bar
length, radius of corotation, and the ratio $\cal R$.  The solid
curves show these quantities for run F, while the dashed curves show
them for run S.  The horizontal dotted line in the bottom panel is
drawn at the value ${\cal R} = 1.4$.}
\label{Rplot}
\end{figure}

A further experiment, shown by the dashed curve in
Figure~\ref{omevol}, has the same numerical parameters as run F.  In
this experiment, we rotated the radius vector to each disk particle
through a random angle before we began, effectively making a different
draw of disk particles from the parent population, which leads to a
different spectrum of initial disk density fluctuations with a similar
amplitude.  In this case, even with the same numerical parameters as
run F, the bar slows in general agreement with the other results.  We
denote this experiment as run S (for slow).

Aside from the anomalous run F, the simulations shown by solid lines
in Figure~\ref{omevol} have numerical parameters that were changed
from those listed in Table~\ref{params} for Runs F \& S.  The
differences were as follows: We both doubled and halved the basic time
step, we increased the resolution of the polar grid, reduced the
spacing of the grid planes, halved the softening parameter (on the
finer grid), both increased and decreased the number of shells in the
spherical grid, and both reduced and increased $l_{\rm max}$ to 2 and
8 on the spherical grid.  Finally, we imposed reflection symmetry
about the disk mid-plane in one of these runs.  In all these runs,
$\Omega_p$ declined at approximately the same rate as in run S.

The solid curves in Figure \ref{Rplot} show the time evolution of the
bar amplitude, bar semi-major axis $a_B$, corotation radius $R_c$, and
the ratio ${\cal R} = R_c/a_B$ for run F.  (Our procedure for making
these measurements from simulations is described in the Appendix.)  The
dashed curves, on the other hand, show the same quantities from run S;
it is clear that the bar in run S soon becomes, and remains,
unacceptably slow.  We find that all other models from
Fig.~\ref{omevol} that slow early show a rapid rise in ${\cal R}$
similar to that in run S.

As reported by VK03 for their bar, the bar in our run F remains
acceptably fast, ${\cal R} \simeq 1.4$, for a long time, although
${\cal R}$ increases when the bar finally begins to slow.  It can be
deduced from Figures 10 \& 14 of VK03 that the corotation radius has
increased from $\sim 6.9\;$kpc to $\sim 10.5\;$kpc, while they report
(their \S 7) that the bar length in their model A$_1$ is 6-6.5~kpc
after 8.5~Gyr of evolution.  Thus the final value of ${\cal R}$ in
their simulation is indeed significantly larger than 1.4.

\begin{figure}[t]
\plotone{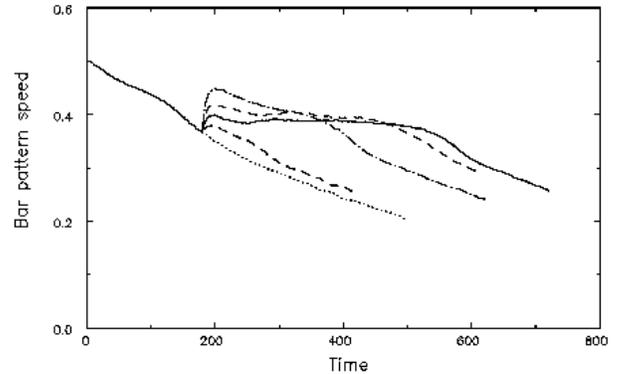}
\caption{\footnotesize The time evolution of $\Omega_p$ of a rigid bar
in an experiment with non-interacting halo particles.  The dotted
curve shows the unforced evolution.  The bar pattern speed is driven
back up between times 180 and 200 to 0.4 (solid line), to 0.38 \& 0.42
(dashed lines) and to 0.45 (dot-dashed line).}
\label{drive}
\end{figure}

\begin{figure*}[t]
\epsscale{1.5}\plotone{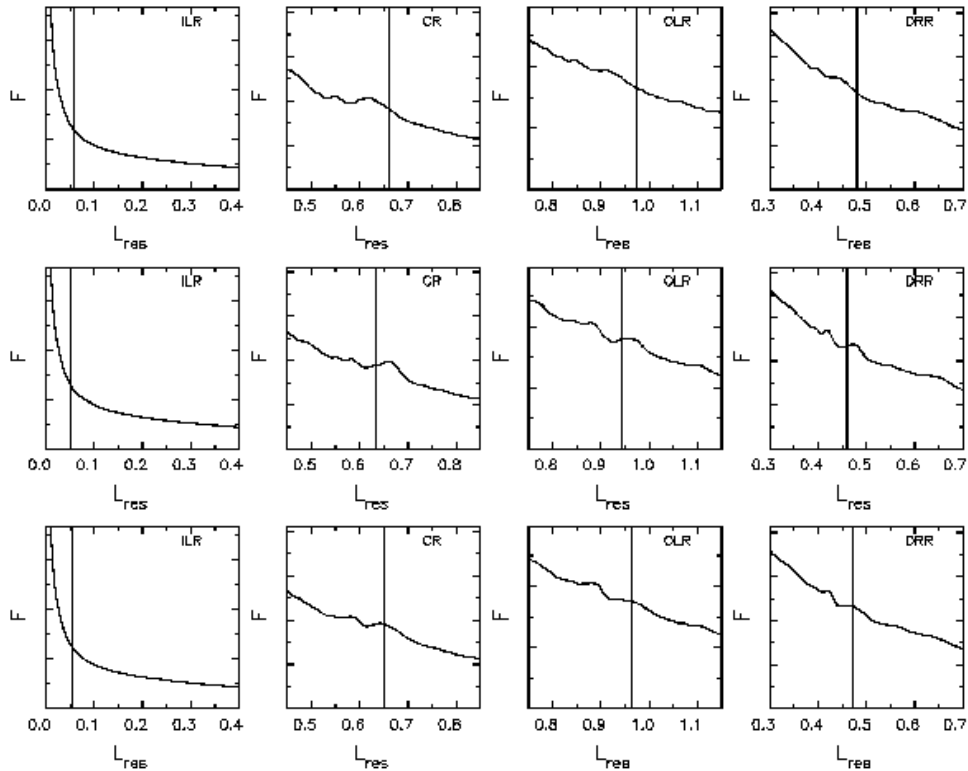}\epsscale{1.0}
\caption{\footnotesize The mean density $F$ of particles as a function
of the resonant angular momentum, $\Lres$, at the four major
resonances at three different times in the model shown by the solid
line in Fig.~\ref{drive}.  The columns of panels are labeled by the
resonances; the top row is for time 176 during the period of normal
friction, the middle row is for time 224 when friction is very weak,
and the bottom row is at time 520 as normal friction starts to resume.
The scaling of $F$ is linear from zero.}
\label{Lsmooth}
\end{figure*}

Thus, our experiments shown in Figure~\ref{omevol} are in excellent
agreement with almost all aspects of the results obtained by VK03
using their entirely different numerical technique, except for the
time at which $\Omega_p$ decreases from $\sim 27\;$km/s/kpc to $<
20\;$km/s/kpc.  In most of our experiments, this happened earlier than
found by VK03, but it occurred later in our anomalous run F.  We even
concur that the bar's corotation circle remains acceptably close to
the bar end in run F for as long as $\Omega_p$ remains $\ga
25\;$km/s/kpc.

We account for our discrepant result in the next section, and that
obtained by VK03 in \S\ref{artifact}.

\section{Metastability}
\label{metastability}
\subsection{Restricted experiment}
Lin \& Tremaine (1983) used restricted $N$-body simulations to show
that driving the system with a perturbation (a companion galaxy in
their case) at a constant frequency, considerably delayed the onset of
friction, once unforced evolution was allowed.  They suspected, but
were unable to show, that the absence of friction was due to all the
resonant particles having been scattered.  Here we show friction can
also be suppressed in the bar case.  At first, we again suspected that
the absence of friction was due to cleared resonances, but this turned
out not to be the case, and we demonstrate that it is due to an
adverse gradient of particle density about the major resonances.

Figure~\ref{drive} shows the evolution of $\Omega_p$ of a rigid bar in
a test-particle halo.  The halo, represented by 10M particles, has a
Hernquist (1990) density profile with an isotropic distribution
function, and the uniform-density, ellipsoidal bar has a mass that is
2\% of the halo mass and an axis ratio $1 : 0.2 : 0.05$, with the
long-axis equal to the scale radius of the Hernquist profile.  (This
is the fiducial experiment described in \S6 of Paper I, where further
details are given.)

The dotted curve shows the evolution of $\Omega_p$, dictated by
conservation of angular momentum for a fixed moment of inertia, as the
bar experiences dynamical friction.  The solid curve shows what
happens when we drive $\Omega_p$ back up to 0.4 between times 180 and
200.  The bar pattern speed, which is free to evolve as a result of
halo friction after $t=200$, stays approximately constant for some
time, and the onset of friction is considerably delayed.  The other
curves show the evolution when $\Omega_p$ is driven back up to
different values.

\subsection{Normal and anomalous gradients}
In Paper I, we introduced the function $F(\Lres)$, which is the
phase-space density reduced to a function of a single variable by
averaging over all orbit phases and eccentricities at fixed $\Lres$.
For any orbit, the frequency difference from resonance is $\Omega_s =
(n\Omega_\phi + k\Omega_r)/m - \Omega_p$, where $\Omega_\phi$ and
$\Omega_r$ are the angular frequencies of the orbit in the unperturbed
spherical potential (\S3.1 of Binney \& Tremaine 1987), and $n$, $k$,
and $m$ are integers.  The quantity, $\Lres$, is the angular momentum
of a circular orbit that is the same distance in frequency,
$\Omega_s$, from the resonance as an orbit of arbitrary eccentricity
-- see Paper I for a fuller description.  The four resonances that are
most important for friction are: the corotation resonance (CR) where
$n=m$ \& $k=0$, the inner and outer Lindblad resonances (ILR and OLR)
where $n=m=2$ \& $k=\mp1$, and the direct radial resonance (DRR) where
$n=0$, $m=2$ \& $k=1$.

Figure \ref{Lsmooth} shows the function $F(\Lres)$ near each of the
four major resonances at three different times in the simulation shown
by the solid line in Fig.~\ref{drive}.  As shown in Paper I, the
distribution of particles about each major resonance is a generally
decreasing function of $\Lres$, but with a shoulder associated with those
resonances that contribute strongly to friction.  The top row is for
$t=176$ during normal friction, the middle row is at $t=224$ during
the metastable phase while the bottom row is at $t=520$ as normal
friction is about to resume.  We first discuss the situation at
corotation (CR, second column): at $t=176$, the distribution has a
local maximum on the low-$\Lres$ side of the resonance as was already
reported in Paper I.  When the pattern speed is driven back up, the
resonance moves to the other side of the local maximum where the
gradient is slightly positive, as shown at $t=224$; finally, as normal
friction resumes ($t=520$) the resonance is just passing the local
maximum.  Similar behavior can also be discerned at the OLR \& DRR.

These figures make it clear that the rise in $\Omega_p$ was enough for
$dF/d\Lres$ to have changed sign at the three resonances that are most
important for friction.  Friction is greatly reduced when the slope of
$F(\Lres)$ becomes positive in the immediate vicinity of the
resonances that dominate the angular momentum exchanges with the bar.
The change in gradient removes the usual excess of gainers over losers
that is responsible for friction.\footnote{The gradient of the usual
distribution function, $\partial f/\partial L$ at constant radial
action, must be nearly flat at the most important resonances in order
that exchanges with the perturbing potential are neutral.  The
positive gradient in $F$ is probably a result of its complicated
relation to $f$.\par} Strong friction does not resume until the
gradients at the dominant resonances become decisively negative once
more.

\subsection{Long-term evolution}
The solid line in Fig.~\ref{drive} is not precisely flat after
$t=200$, and a slow decrease in $\Omega_p$ is discernible.  The mild
friction that causes the slow decrease probably results from exchanges
at higher-order resonances.  The set of possible resonances between a
halo orbit and a rotating perturber is large, since a resonance arises
for any set of integers, $k, \, m, \, n$.  Thus even if the important
low-order resonances cause no friction, some residual friction is
provided by exchanges at the many higher-order resonances; gradients
in $F(\Lres)$ at these weak resonances were not much affected by the
previous evolution.  The weaker coupling between the
perturbation and particles at higher-order resonances (Hernquist \&
Weinberg 1992; Paper I) leads to slower, but non-zero, angular
momentum transfer.  (We have been unable to find convincing evidence
of angular momentum exchanges at a number of possible resonances,
which is hardly surprising as friction is so weak.)  Thus, the pattern
speed gradually decreases until the full friction force picks up again
when gradients of $F(\Lres)$ at the dominant resonances are negative
once more.

The lower dashed line in Fig.~\ref{drive} shows friction was not
greatly affected when the rise of $\Omega_p$ was insufficient to move
the CR past the maximum of $F(\Lres)$.  The upper dashed line, on the
other hand, shows that friction reappears briefly after $\Omega_p$ is
driven up to a slightly larger value, but again becomes anomalously
weak for a long period.  Analysis of $F(\Lres)$ in this case shows
that the slight initial drop in pattern speed allows the resonances to
reach the adverse gradient in $F(\Lres)$.  The duration of the
metastable phase is slightly shorter because the local maximum in
$F(\Lres)$ is eroded somewhat by the evolution after the pattern speed
is driven up.  After the pattern speed is driven up to a still higher
value, the bar slows continuously with only a slight reduction in
friction as the bar passes through $\Omega_p \approx 0.38$; \ie\ too
large a rise in $\Omega_p$ does not lead to metastability.

\begin{figure}[t]
\epsscale{0.8}\centerline{\plotone{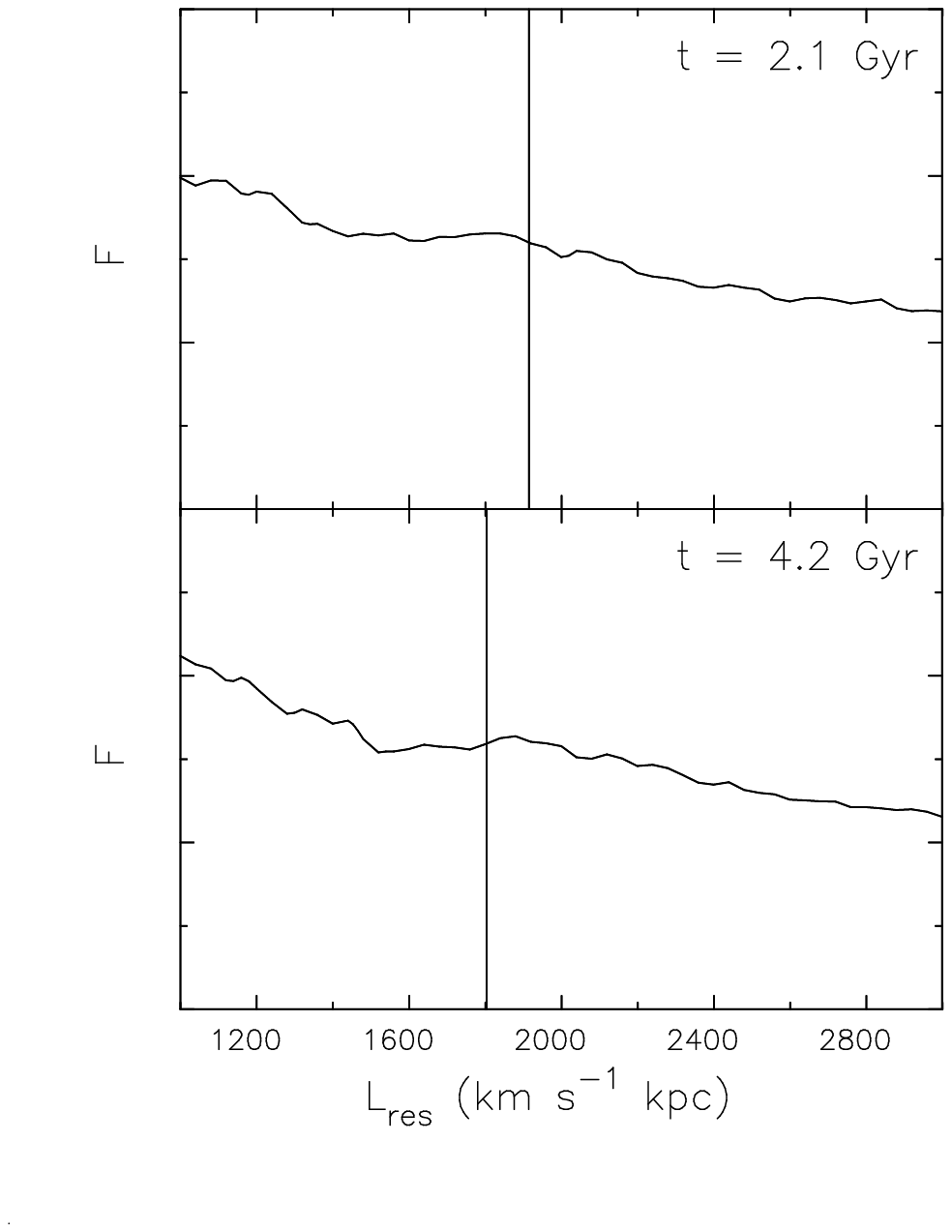}}\epsscale{1.0}
\caption{\footnotesize The function $F(\Lres)$ at corotation at two
different times in run F from Fig.~\ref{omevol}.  The earlier time,
$t=2.1$, is during normal friction, while $t=4.2$ is after the pattern
speed has risen.}
\label{LsVK}
\end{figure}

\subsection{Relevance to Figure 1}
It therefore seems likely that the long period of steady bar rotation
reported by VK03, and the even longer such period in our run F, were
caused by the upward fluctuations in $\Omega_p$ in both cases visible
in Fig.~\ref{omevol}.  These upward fluctuations must have raised
$\Omega_p$ to values at which the gradients in $F(\Lres)$ at the
dominant resonances are reversed.

Figure \ref{LsVK} shows this to be the case in our run F.  In order to
obtain this figure, we needed to estimate $\Lres$ for each particle in
this fully self-consistent simulation with a disk and bar.  Unlike in
the restricted experiments just reported, the mass profiles of the
disk and halo evolve continuously because a bar forms in the disk and
the halo was not initially very close to equilibrium.  We therefore
approximated the potential well in the simulation as a time averaged,
spherical mass profile, extracted from the simulation over a short
interval.  We then computed the two frequencies of every halo particle
from its energy and angular momentum defined by its instantaneous
position and velocity at a moment in the middle of the period of the
time averaging.

The features in Fig.~\ref{LsVK} are not quite as clear as those in
Fig.~\ref{Lsmooth}, possibly because we had to erase the non-spherical
components of the disk, bar, and halo, in order to compute the orbit
frequencies.  (We had to discard a tiny fraction of particles that had
less energy than that of a circular orbit of their angular momentum in
this approximate potential.)  Nevertheless, a feature that resembles
the shoulder at corotation in Fig.~\ref{Lsmooth} can be seen, and the
rise in pattern speed between the two times moves the resonance across
the shoulder to change the gradient in $F(\Lres)$.

We can only speculate why there was a longer delay before full
friction resumed in our run F than in the experiment reported by VK03.
One possible reason might be that the upward fluctuation in their run
may have taken $\Omega_p$ only just past the local maximum of
$F(\Lres)$.

It is also possible that collisional relaxation in simulations with
self-gravity allows halo orbits to diffuse slowly in phase space,
thereby eroding the local maximum in $F(\Lres)$ (which could not have
happened in the restricted experiments of the previous subsection).
This effect seems to be minor in our run F, since the higher-order
resonance explanation accounts for the resumption of friction on the
appropriate timescale (compare Figs.~\ref{omevol} \& \ref{drive}).  In
fact, if the rapid orbit diffusion worry raised by Weinberg \& Katz
(2002, 2005) were important, the metastable state could not persist
for long.  (See also \S \ref{collisionality}.)

Note that the absence of friction is not simply a question of
inadequate coverage of phase space by the finite number of particles,
since the bar still slows as normal (the dotted curve in
Fig.~\ref{omevol}) when the density of particles is reduced by a
factor of 10.  (See also Paper I.)

\subsection{What happened in run F?}
We have attempted to understand what caused the upward fluctuation in
our run F.  Detailed comparison of this run with others that slowed
has led us to conclude that it was caused by a random interaction of
the bar and a spiral pattern in the disk that occurred at a rare
relative phase so as to add angular momentum to the bar -- spirals
generally have the opposite effect at most relative phases.

It should be noted that most spiral activity in disk simulations is
genuinely stochastic.  As the initial spiral patterns are determined
by the swing-amplified (Toomre 1981) seed spectrum of particle noise
laid down by the random selection of particles, the very first
features are insensitive to parameters such as grid resolution.
However, the subsequent patterns can differ macroscopically as the
result of seemingly insignificant changes to numerical parameters.
Slight changes to the disk responsiveness or central attraction cause
small differences to the early evolution, but the subsequent sequence
of spiral patterns in models that differ in this way visibly diverges
in remarkably few dynamical times.

These differences are largely responsible for the scatter in slow-down
times (but not rates) already noted in Fig.~\ref{omevol}.
Fortunately, the statistical properties of the disk at later times are
not sensitive to these differences, and the divergent sequences of
spirals result in similar disk random motion and -- excepting run F --
bar speed evolution.

\vfill\eject
\section{Adaptive refinement}
\label{artifact}
The evidence presented so far indicates that stochastic disk evolution
can occasionally, with our code, put the bar into the metastable
state, which was also reached in the simulation reported by VK03.  It
is possible they were just unlucky to find this state, although they
report little bar friction over the (short) period of evolution in
their model B also, making the chance explanation seem unlikely.

The principal numerical difference between their method and ours is
their use of adaptive mesh refinement (Kravtsov \etal\ 1997) and we
now show that this feature may make the metastable state easier to
reach.  Our reasoning is as follows: As a bar develops, particles
become more closely packed in the bar than they were in the disk and
an adaptive code will cause short-range gravitational forces in the
bar to strengthen over those pertaining if resolution/softening were
fixed.\footnote{The forces between particles in grid codes have a
softened form inside $\sim 3\;$mesh spaces (see \eg\ Fig.~14 of
Sellwood \& Merritt 1994).\par}  Stronger forces, or steeper potential
gradients towards the center, have two effects: they cause the bar to
contract, and they cause the orbital periods of particles at the same
mean galacto-centric distance to be somewhat shorter.  Both effects
will raise the pattern speed of the bar over what it would have been
in a fixed resolution code, which could possibly be sufficient to push
the bar into the metastable state.

\begin{figure}[t]
\plotone{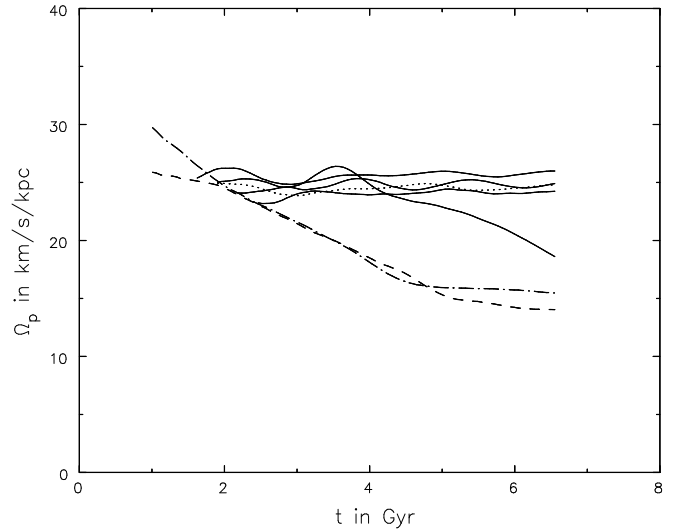}
\caption{\footnotesize The dashed line shows the time evolution of
$\Omega_p$ in a fully self-consistent simulation with 4 times larger
$\epsilon$ and finer grid than were used for run F.  The dot-dash line
shows the evolution with the same softening length as run F, but run
on the finer grid.  The solid lines show results when $\epsilon$ is
reduced to the lower value after some evolution with the larger
softening.  The dotted curve shows the effect of only halving
$\epsilon$.}
\label{adapt}
\end{figure}

\subsection{Mimicry}
\label{mimicry}
To test this hypothesis, we ran a suite of further experiments with
model A$_1$ from VK03, using different softening lengths, $\epsilon$,
and higher grid resolution.\footnote{Softening length and grid
resolution can be varied independently in the convolution method used
for our cylindrical polar grid.
Changes to $\epsilon$ have almost no effect, however, unless the grid
cell dimensions are $\la\epsilon$, which, for a polar grid, is
possible in the inner parts only.  The softening length, $\epsilon$ is
larger than the $(R,\phi)$ cell dimensions out to $R\simeq 0.3 R_{\rm
d}$ for the shortest value of $\epsilon$ used in Fig.~\ref{adapt}, and
this radius increases nearly linearly with $\epsilon$ for a fixed grid
size.\par} The bar slows normally in the basic model of this suite,
shown by the dashed line in Figure~\ref{adapt}.  The softening length
in this case, $\epsilon = 0.05R_d = 175\;$pc, is 4 times greater than
for most of those in Fig.~\ref{omevol}, and for the run shown by the
dot-dashed curve in this Fig.~\ref{adapt} that also slows normally.

But the bar stayed fast, at least for a while, in the four cases shown
by solid lines when softening was reduced by a factor of 4 after
different periods of evolution with the larger $\epsilon$.  In each of
these four cases, softening was decreased abruptly at one instant in the
interval $2 \la t \la 2.8\;$Gyr, \ie\ during the later period of bar
growth.  (In one of these four cases, the bar speed experiences a
second upward fluctuation and then declines; we comment on that case
in \S6.) The curve shown by the dotted line is for another case in
which softening was merely halved -- again the bar stayed fast.

This crude stratagem is intended to mimic the effects of adaptive
refinement, although in our code the change is uniform over the inner
cylindrical grid and abrupt.  The change in softening affects only the
small fraction of disk particles in the inner grid; the polar grid has
cell sizes that increase with cylindrical radius and inter-particle
forces are soon limited more by the grid than by the formal softening.
It should be emphasized that this change does not affect forces on
halo particles, which are computed from the spherical grid.  As a
result, the abrupt change in softening has a barely noticeable effect
on the equilibrium of the model.

The surface density along the ridge of the inner part of the bar is
almost 4 times higher than the original axisymmetric disk (Fig.\ 15 of
VK03) well after the bar has formed.  (We find a somewhat larger
increase that relaxes back later as the bar becomes slightly rounder.)
Since this surface density change takes place before the bar thickens
through buckling instabilities, the mean distance between disk
particles in this region decreases by about a factor 2 as the bar
forms.  Thus a reduction of softening by a factor of two in our code
is approximately equivalent to adaptive refinement in the ART code
employed by VK03 for the same model.  The effect of the change in
softening strongly suggests that the metastable state that appears in
several of the experiments reported by VK03 is a numerical artifact of
adaptive refinement.  Their finding that bars slow as normal if they
limit the level of mesh refinement (their \S7) reinforces this
suspicion.

VK03 report an additional model, their model C, which had more
particles and the highest resolution of their adaptive scheme was
limited to 100 pc, as opposed to 44 pc for their other simulations.
It is therefore unlikely that the pattern speed in this simulation
would rise due the numerical artifact that affected model A$_1$, and
possibly also their model B.  Unfortunately, they did not run their
simulation of model C for very long; the bar pattern speed is
declining over the second half of the evolution, and the bar would
probably have become slow had they continued the calculation.

\vfill\eject
\subsection{Discussion of softening}
Adjusting the softening parameter, or the grid resolution, dynamically
clearly changes the behavior.  Why should this be?

Softening can be thought of as an operation first to convolve the
particle distribution with a smoothing kernel, and then to solve for
the full Newtonian field of the smoothed density distribution.
Ideally, the softening length, $\epsilon$, should be set large enough
that many particles lie within one softening volume, but short enough
so as not to smooth, or bias, gravitational potential gradients.

If $\epsilon$ is large compared to the mean separation of particles
in a region of quasi-uniform density,
then small changes to the width of the softening kernel will not alter
the smoothed density, and the global gravitational potential will be
unaffected.  Conceptually, we could imagine increasing $N$ as
$\epsilon$ is reduced all the way to the double limit $N \rightarrow
\infty$ as $\epsilon \rightarrow 0$, without affecting the potential
at all, but only if $\epsilon$ starts small enough not to smooth any
density gradient significantly.  Thus, wherever the twin ideals of
many particles per softening volume and a softening length shorter
than any density gradient hold, adaptive refinement would not change
the dynamics.

Unfortunately, these ideals are generally not achievable in practical
simulations of realistic stellar systems.  The desire to resolve steep
density gradients causes simulators to prefer values of $\epsilon$ on
the low-side, in order to maximize spatial resolution, or to minimize
bias, which leads to few particles per softening volume; for example,
VK03 continue to sub-divide any cell that contains more than 4
particles.  The penalty for this preference is an increase in the
variance in the gravitational potential, but the increase in
relaxation rate, through the change to the Coulomb logarithm, is quite
small.

However, global gravitational potential gradients in the disk plane
are softening dependent unless $\epsilon \ll z_0$, the vertical
density scale height.  Density gradients in the $x$- and
$y$-directions are quite shallow, even in a bar, but vertical
gradients in thin disks are a different matter.  It is hard to ensure
$\epsilon \ll z_0$ for a thin disk in any code.

The consequences of inadequate vertical resolution are two-fold.
First, weakened vertical forces increase the vertical oscillation
period of disk particles.  Such a bias has a small impact on the
in-plane motion of particles, which are mostly decoupled from their
vertical motions, {\it except} at later times when the bar buckles.
Second, in-plane potential gradients are weakened by softening (see
\eg\ problem 6-5 of Binney \& Tremaine 1987) unless $\epsilon \ll z_0$. 
Thus, the dynamics of bar formation is different in simulations with
different, but fixed, softening lengths.

Initially, $z_0 \approx 140\;$pc in model A$_1$ (not 250~pc as stated
in Table 1 of VK03) and the disk does not thicken much until some time
after the bar has formed.  The effect of using a different, but
unchanging, softening length in our code is that the bar forms a
little earlier and with a higher $\Omega_p$ when the softening length
is shorter.  Steady bar friction results in both cases shown in
Fig.~\ref{adapt}, braking the bars at similar rates.  This mild
dependence on softening length could be avoided only if $\epsilon \ll
z_0$, and $N$ were increased appropriately -- an ideal that is
numerically expensive for a thin disk.

The ART code used by VK03 for this model refined cells to a size of
22pc, or an effective minimum softening of $\sim 40\;$pc, but it was
probably larger over most of the disk.  Adaptive refinement must
therefore increase the in-plane forces as the bar strengthens, causing
the bar to speed up purely for this numerical reason.  Thus adaptive
refinement may trigger the metastable state.

\begin{figure}[t]
\plotone{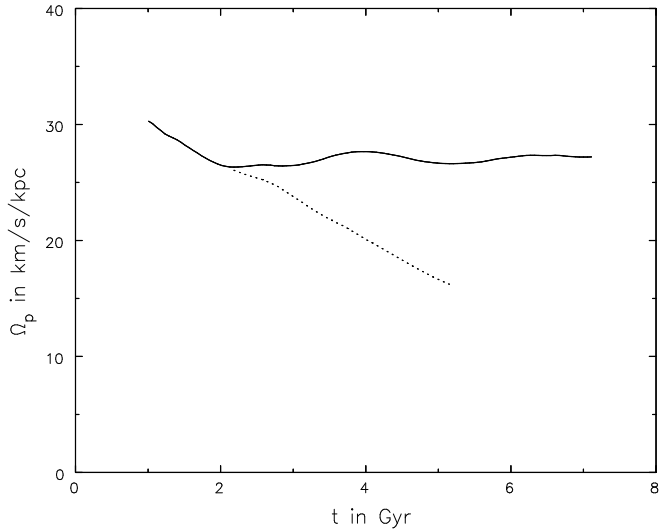}
\caption{\footnotesize The solid line shows the time evolution of
$\Omega_p$ in a fully self-consistent simulation in which we increased
the masses of the 1\% most bound particles by 50\% at $t=2.24\;$Gyr.
The dotted curve shows the result without this change.}
\label{inflow}
\end{figure}

\section{Metastability in real galaxies?}
\label{gasinflow}
We have shown that long periods without friction can occur in purely
collisionless $N$-body simulations if the bar pattern speed rises for
some reason.  This can happen occasionally at fixed softening, or
spatial resolution, but it is more likely to happen when spatial
resolution is adaptive; in any case, the expected braking is recovered
when the numerical procedure or parameters are changed.

It is interesting to ask whether real bars could avoid friction in a
similar manner.  It seems possible that random spiral events, or even
the interaction with a minor satellite galaxy, could occasionally
increase $\Omega_p$ and allow strong braking to be avoided for a
while; but such behavior seems likely to be the exception, rather than
the rule.  However, real galaxies contain gas, which behaves
differently from stars; gas is widely recognized to flow in towards
the center of a bar.  Large gas accumulations are found in regions a
few hundred parsecs across (\eg\ Sakamoto \etal\ 1999) which are
sometimes resolved as inner gas rings.  While the gas accumulations
are a small fraction of the total disk mass, it is possible the mass
increase in the center could well be enough to raise $\Omega_p$.

To test this, we took a model in which the bar was slowing as normal,
and increased the masses of each of the 1\% most bound disk particles by
50\% at $t = 2.24\;$Gyr -- \ie, we increased the total disk mass by 0.5\%,
by adding mass at the bar center only.  Figure~\ref{inflow} shows that
frictional braking ceased and $\Omega_p$ stopped declining.

The central mass added in this experiment, $\sim 2 \times
10^8\;\hbox{M}_\odot$, is quite consistent with observed gas masses
in bar centers.  It therefore seems entirely possible that gas inflow
in bars could cause a bar to speed up enough to turn off dynamical
friction even in a dense halo.

\begin{figure}[t]
\plotone{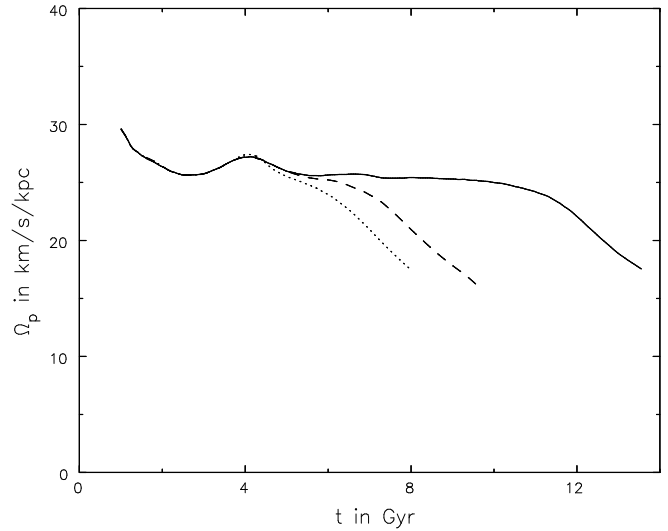}
\caption{\footnotesize The solid line shows the time evolution of
$\Omega_p$ in run F, again reproduced from Fig.~\ref{omevol}.  The
dashed curve shows $\Omega_p$ when a 1\% mass satellite galaxy flies
by, passing a peri-galactic distance of 30~kpc at $t=4.2\;$Gyr.  The
bar is jolted out of the metastable state and begins to slow.  The
dotted curve shows the bar slows after the disk is discontinuously
rotated forwards relative to the halo by $10^\circ$ at $t=4.2\;$Gyr.}
\label{jolts}
\end{figure}

\section{Fragility of the metastable state}
\label{fragile}
If the bar in an isolated galaxy with a dense halo has been spun up so
that it experiences little friction, our simulations suggest that it
might be several Gyr before the higher order resonances can reduce the
pattern speed to the point at which strong friction resumes.

However, real galaxies are not isolated, and are subject to numerous
small perturbations from infalling dwarf galaxies as well as the
passage of larger galaxies at greater distance; \eg, the Milky Way
today has the Sagittarius dwarf, the Magellanic Clouds, and other
dwarf companions.  In addition, the $\Lambda$CDM model of galaxy
formation predicts a much larger number of dark ``mini-halos'' (Moore
\etal\ 1998; Klypin \etal\ 1999) orbiting within the halo of a large
galaxy.  Since the absence of friction depends on a pre-arranged
distribution of particles at the major resonances, the constant
stirring of a galaxy halo by these perturbations might well cause
strong friction to resume much earlier.  We have conducted a number of
experiments in order to investigate this possibility.

We perturbed our run F at $t=4.2\;$Gyr, which is in the metastable
state, by a small satellite galaxy that flies by.  We model the
satellite as a rigid Plummer sphere with a scale radius equal to the
exponential disk scale.  We introduce it at a distance of 60~kpc from
the center of the galaxy at $t=4.2\;$Gyr, when its mass is very small,
and increase its mass gradually to its final value at $t=4.5\;$Gyr.
It has a mildly hyperbolic, polar orbit that crosses the disk plane at
a radius of some 30~kpc at about $t=4.9\;$Gyr.  In order to ensure
that the main galaxy remains optimally centered in the grid, we shift
the grid center every 16 time steps to coincide with the densest point
(a feature that was enabled in the majority of the simulations already
reported); all three components of total linear momentum are well
conserved after the intruder has reached full mass.  We have tried
satellite masses that are 1\%, 2\% and 5\% of the total (disk + halo)
mass of the main galaxy.  Even for the 1\% mass satellite, the halo is
sufficiently perturbed for bar friction to resume as shown by the
dashed curve in Fig.~\ref{jolts}; the solid line again shows the
evolution of $\Omega_p$ in run F.

In a separate experiment, we perturbed run F at the same moment by
rotating every disk particle forwards through ten degrees and then
allowed the system to evolve freely once more.  The purpose of this
exercise is instantly to change the orientation of the bar relative to
the density response in the halo, and therefore to change the orbital
phases of the halo particles relative to the bar.  While artificial,
the sudden change of bar phase could perhaps resemble the effect of
interaction with a strong spiral pattern.\footnote{The outer disk in
our simulation supports generally rather weak spiral patterns after
the bar has formed.  Barred galaxies having gas and on-going star
formation would be expected to have stronger spirals.}  Just such an
event seemed to be responsible for the upward fluctuation followed by
early slow-down of one of the bars in Fig.~\ref{adapt}.  The dotted
curve in Fig.~\ref{jolts} shows the evolution of this perturbed model;
the upward fluctuation in $\Omega_p$ immediately after the imposed
change is an artifact caused by our measuring the bar pattern speed
from the slope of its phase angle with time, which includes our
imposed $10^\circ$ discontinuous change.  Once this feature is outside
our fitting window, $\Omega_p$ decreases at the usual rate.  Thus
jolting the system in this manner again tipped it out of the
metastable state.

The results of both experiments indicate that extremely mild
disturbances to the system are sufficient to cause friction to resume.
We conclude that the metastable state is highly fragile and it is
unlikely it could persist in real galaxies for long.

\section{Further numerical issues}
\subsection{Spatial resolution}
The spatial resolution of the fixed Cartesian grid used in our earlier
experiments (DS98, DS00) was indeed quite low and VK03 suggest that it
was inadequate to capture the correct physics.  We disagree.

As stated by Hernquist \& Weinberg (1992) and shown in Paper I,
dynamical friction is dominated by the quadrupole field of the bar,
which couples most strongly to the low order resonances.  A good
approximation to the correct quadrupole field can be obtained with a
mesh of quite low spatial resolution, and the torque between a given
bar and halo should be little affected by spatial resolution.

We have already presented supporting evidence for this statement,
since we have shown that the rate of slow-down of the bar is
insensitive to softening length and grid resolution.  The different
runs shown in Fig.~\ref{omevol} have a range of softening lengths and
grid resolutions and $\epsilon$ differs by a factor 4 between the two
runs with unchanging $\epsilon$ shown in Fig.~\ref{adapt}, which track
each other remarkably closely.  Furthermore, once friction picks up in
the simulation by VK03 (the dot-dash curve in Fig.~\ref{omevol}), the
rate of slow down is quite comparable to that we find in our
simulations.

\begin{figure}[t]
\plotone{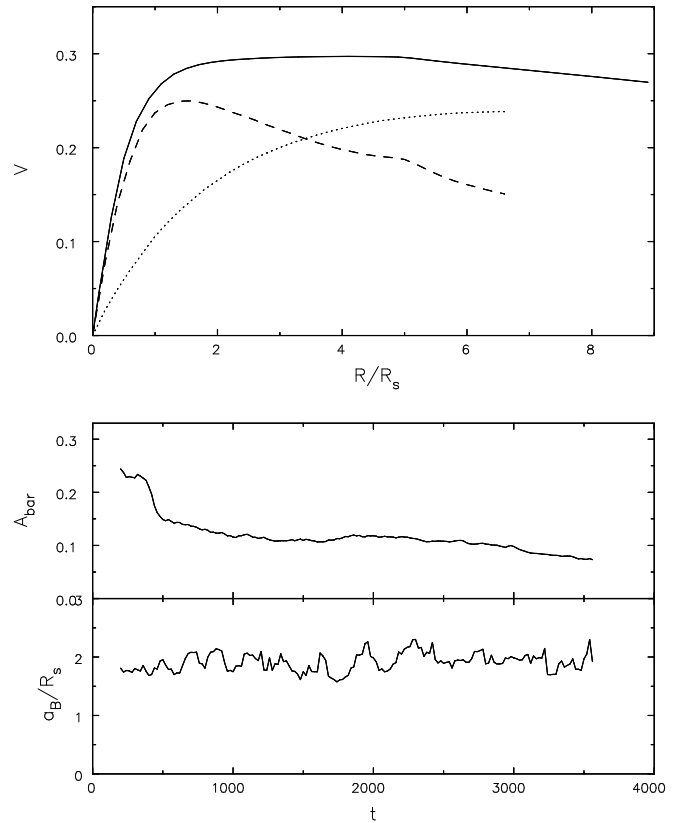}
\caption{\footnotesize (a) The rotation curve (solid line) and
separate contributions of disk (dashed) and halo (dotted) for a
maximum disk model, similar to that reported in DS00.  (b) The bar
amplitude and length as a function of time.  Note that this initially
large bar neither grows in length nor amplitude.}
\label{maxdisk}
\end{figure}

\subsection{Bar size}
The pattern speed and spatial scale of the bar instability, and the
corresponding properties of the resulting bar, depend on many factors.
Two of the most important are the steepness of the inner rotation
curve and the velocity dispersion of the disk.

VK03 suggest that inadequate spatial resolution can lead to longer
bars.  Indeed, if the central density gradient (of disk, bulge and
halo particles) is steep, inadequate spatial resolution will weaken
the central forces and this numerical bias leads to a rotation curve
that rises more slowly than it should.  Since more slowly rising
rotation curves lead to longer bars (Sellwood 1981) with lower pattern
speeds, the size of the bar may be artificially enhanced if inadequate
spatial resolution smoothes the sharper forces expected from a steep
density gradient.  

However, smaller bars are not the inevitable consequence of improved
spatial resolution.  The large size of the bars in DS00 was a
consequence of the shallow density gradients in our models, which we
deliberately selected because we were fully aware that the code we
were using was unable to resolve steep gradients.
Figure~\ref{maxdisk} shows the rotation curve and bar evolution of a
model deliberately set up to have a halo with a large, low-density
core to resemble the ``maximum disk'' model reported in DS00, but
evolved with our newer high-resolution code.  As the similar model in
that paper, the disk has a Kuzmin-Toomre density profile with scale
radius $R_s$, thickness $0.05R_s$ and initial velocity dispersion to
make $Q=1.0$ at all radii; the low-concentration halo has a polytropic
DF, a mass about 5 times the disk mass, and extends to $r=28R_s$.  The
bar that forms in our high-resolution, hybrid-grid code is
similar in size and amplitude to that reported in DS00 for the
equivalent model run on their coarse grid, as shown in the lower
panels in Fig.~\ref{maxdisk}.  Thus bar size need not be related to
spatial resolution.

DS00 compared the strengths and sizes of the bars in their models with
bar properties determined from NGC~936.  They concluded that the bars
in their maximum disk models were quite comparable to strong bars in
real galaxies.  Note also that the bar length in Fig.~\ref{maxdisk} is
reckoned relative to the length scale, $R_s$, of the initial KT disk;
after the bar has formed, the outer disk is characterized by an
exponential profile with a scale length about $2.5R_s$.  This
important difference has already been noted by VK03.

However, bars become longer and stronger when subject to strong frictional
braking as shown in Fig.~\ref{Rplot} and previously by DS00 and
Athanassoula (2002).  DS00 showed that bars that had been braked in
denser halos were significantly stronger than that in
NGC~936, which might be typical of strongly barred galaxies.  Even the
mild braking in their ``maximum disk'' model strengthened the bar to
the point where it was only marginally consistent with the data on
NGC~936.  We therefore agree with VK03 that bars often become quite
unrealistically large and strong.  The more realistic bars in models
with maximum disks provide further evidence against the hypothesis
that halos have high central densities: not only does friction
cause bars to become too slow, they also become unrealistically large
and strong.

\begin{figure}[t]
\plotone{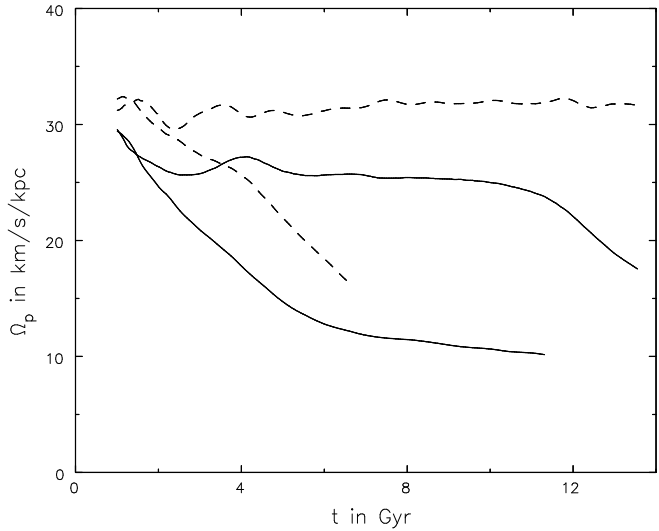}
\caption{\footnotesize The solid lines show the time evolution of runs
F and S, reproduced from Fig.~\ref{omevol}.  The dashed curves show
results when the gravitational forces are restricted sectoral
harmonics $0 \leq m \leq 4 $.}
\label{restrict}
\end{figure}

\subsection{Disk velocity dispersion}
A cooler disk leads to a bar of higher pattern speed that may also
have to be shorter to fit inside its corotation circle.\footnote{A
weak bar may have a lower $\Omega_p$ and end well inside its
corotation circle.}  The trend in $\Omega_p$ is expected from bar
instability theory (Kalnajs 1977), and agrees with previous findings
(Athanassoula \& Sellwood 1986) and with the difference between the
models A$_1$ and A$_2$ reported by VK03.

Figure~\ref{restrict} presents an additional illustration of the
effect of lower velocity dispersion at the time the bar formed; $Q
\simeq 1.5$ in the disks of the models with dashed lines, whereas $Q
\simeq 1.8$ for the models with solid lines.  The solid lines are from
our runs F \& S, reproduced from Fig.~\ref{omevol}, in which sectoral
harmonics $0 \leq m \leq 8$ contributed to the forces computed from
the disk particles.  The dashed lines show results from models in
which all sectoral harmonics $m>4$ were eliminated from the force
determination -- \ie\ a calculation with more heavily smoothed forces.

The disk is cooler in the models with the more restricted forces
because the early stages of spiral evolution are weaker; this is not
to say that strong spiral patterns with more than four arms develop
when more harmonics are included, simply that the patterns that do
develop have steeper density gradients than simple sinusoidal density
profiles, leading to stronger scattering by the density fluctuations.

The pattern speed clearly stayed fast in one of these two models for
as long as we ran it, but a minor change to the numerical parameters
(we increased the number of radial points of the spherical grid) again
caused the bar to slow down.  Notice that $\Omega_p$ slowed at about
the same rate as for the runs with a larger number of sectoral
harmonics; the bar with the higher $\Omega_p$ is somewhat shorter at
first, but grows to about the same size as it slows causing the
torques to become more nearly equal.

\begin{figure}[t]
\plotone{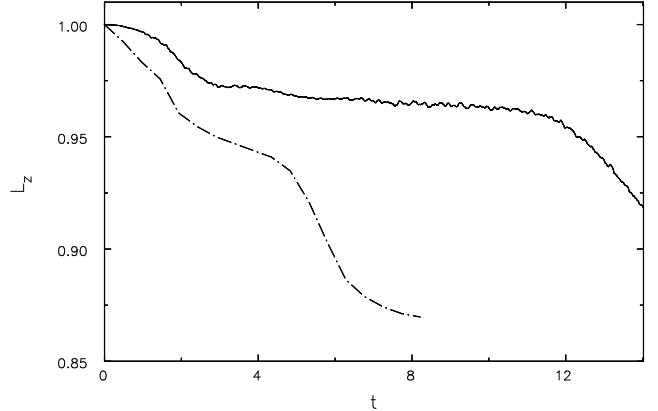}
\caption{\footnotesize The solid line reports the disk angular
momentum in our run F, relative to its initial value, as a function of
time, showing that the torque is weak while the bar rotates
steadily.  The dot-dash line, which is reproduced from Fig.~10 of
VK03, shows the same quantity in their simulation.}
\label{notorque}
\end{figure}

\subsection{Collisionality}
\label{collisionality}
Figure \ref{notorque} compares the angular momentum transferred from
the disk to the halo in our run F (solid curve) with that reported by
VK03 for the same model; the dot-dash line is reproduced from their
Fig.~10.  The disk in the simulation with their ART code loses angular
momentum more rapidly than in ours.  In both simulations, the torque
is strong when $\Omega_p$ is changing, and is weaker while $\Omega_p$
is approximately constant.  However, VK03 find that the disk in their
simulation starts to lose angular momentum right from the outset,
before the bar has formed, and the gradient is again steeper than in
our case while the bar rotates steadily.

The more rapid angular momentum transfer in their model during the
period of steady bar rotation appears to be unrelated to the usual bar
friction.  As already noted in \S\ref{results}, the properties of the
initial bar in their simulation are similar to those in all the
reruns reported here.  We cannot compare our bar strengths with
theirs, since we employ a different measure of bar strength, but
initially similar bars in our simulations (Fig.~\ref{Rplot}) slow both
more and less rapidly than in theirs, making it unlikely that the
difference could be attributed to a stronger bar in their case.

It is possible that the stronger drag between the disk and halo in
their simulation is caused by two-body encounters in their code.  As a
halo particle pursues its orbit, the density of nearby particles peaks
up briefly as it passes through the disk.  In an adaptive code that
refines any grid cell that contains more than just a few particles,
the likelihood of a strong deflection from a close encounter with
another particle is greatly enhanced within the disk.  Since the disk
particles generally have a larger $L_z$ than those in the halo,
deflections caused by close encounters between them will, on average,
transfer angular momentum from the disk to the halo.

VK03 report a measurement of the collision rate in their code in a model
without a disk (their \S3.4), which is an ensemble average of all the
encounters in high- and low-density regions.  Their measurement is
dominated by the cumulative effect of distant encounters.

Here we appeal exclusively to close encounters that produce the
occasional strong deflection.  It is easy to show that, to first order,
deflections are just as likely to add angular momentum to disk particles
as they are to remove it, and a net torque from collisions arises only
to second order.  Since the mean angular momentum exchanged increases
as the square of the angle of deflection, the net collisional torque is
dominated by rare, large-angle deflections.

We suspect that the non-zero disk-halo torque reported by VK03 from the
outset of their experiments results from 2-body encounters as halo
particles pass through the disk.  Note that the density of particles in
the disk mid-plane is more than twice that of the halo alone over a
wide swath of the disk, further increasing the likelihood of a strong
deflection in this critical part of the orbit of a halo particle.  Note
also that collisional encounters cause angular momentum to be lost even
when the disk is axisymmetric and, since they remove angular momentum
from all over the disk and not just from the bar, the bar can rotate
steadily while angular momentum continues to be lost to the halo.  Of
course, encounters must continue to scatter halo particles after the
disk has developed a strong bar, adding to the torque that arises from
dynamical friction.

Such encounters must also scatter the disk particles, and we therefore
expect the disk to thicken more in their experiments than in ours.
Their paper does not provide any information about disk thickness with
which we might compare, however.

Collisionality could be an additional reason that the metastable state
is short-lived in their simulation.  Collisional encounters, whether
between halo-halo particles, or halo-disk particles, would scatter
particles and erode unusual density gradients near resonances,
allowing friction to resume after less evolution than in a more nearly
collisionless code.

\section{Conclusions}
The bars in two simulations reported by Valenzuela \& Klypin (2003),
and those in two of our own reported here, rotate in dense halos for
long periods without the expected friction.  We have shown that this
anomalous behavior arises because the halos lack the decreasing
density of halo particles with angular momentum about the principal
resonances usually responsible for friction.  An inflexion in the
angular momentum density of particles is set up during a period of
normal friction as the pattern speed decreases in the usual manner.  A
subsequent increase in the bar pattern speed can move the resonance
into the region where the local gradient is anomalous with the result
that friction is suppressed for a long period.  Because it relies on
local minima in the distribution of halo particles, we describe almost
frictionless bar rotation as a metastable state.  It does not last
indefinitely, although it can persist for cosmologically interesting
time periods in isolated galaxy simulations.

We argue that the long period of steady bar rotation reported by VK03
in the evolution of their models A$_1$ and B occurred for this reason.
We occasionally find bars enter a metastable state in our own
simulations, but we have shown that changes to the numerical
procedure, or to numerical parameters, can cause or prevent the state
from appearing.  It arises in our models from a rare interaction
between the bar and a stochastic spiral in the outer disk that gives
the bar some angular momentum.  Such an event could also have happened
in the simulations by VK03, but the metastable state is more likely a
result of their numerical method.  We show that the pattern speed is
driven upwards by increases in the central attraction as the grid is
refined in the adaptive scheme used by VK03, making the metastable
state accessible for numerical reasons.  It should be noted that VK03
report in their \S7 that bars slow normally in their code when
adaptive refinement is turned off.  Their model C, in which the level
of refinement was also limited, was not run for long enough to show
the full extent of bar slow-down.

We have shown that when this artifact is avoided, the bar in their
model A$_1$ experiences strong friction and the corotation radius
quickly moves out to an unacceptable distance, relative to the bar
length.  Thus their model is, in fact, consistent with the conclusions
of Debattista \& Sellwood (2000) that strong bars cannot stay fast if
the halo density is high.

We emphasize that bars can rotate at nearly constant pattern speed for
reasons other than metastability.  Friction will be mild, and bars
will not slow much, if either the halo density is low, or if the
bar is weak.  The metastable state describes only strong bars that are
able to rotate with little friction in dense halos because the
distribution of resonant particles that would normally give rise to
strong friction has acquired an anomalous gradient in angular momentum.

If strong bars in real galaxies could rotate rapidly in dense halos
for long periods, the constraint on halo density proposed by us (DS98,
DS00) would be severely weakened.  In simulations without dissipation
such behavior is a numerical artifact, but we have shown that
mimicking a realistic amount of gas inflow in the bar can indeed cause
the bar to speed up sufficiently for friction to stop, suggesting that
the metastable state could arise quite naturally.  However, we have
also shown that the metastable state is highly fragile, and does not
survive even minor internal, or external perturbations.  We find
that a fly-by satellite having as little as 1\% of the main galaxy
mass provides a sufficient disturbance to the system that normal
friction resumes.  It is therefore unlikely that many galaxies
could survive in the metastable state for long periods.  It is doubly
unlikely that SB0 galaxies, the majority of galaxies for which $\cal
R$ is measured, are in this state, since they lack gas that can be
driven into the center to raise the bar pattern speed.

\acknowledgments
\noindent We thank Anatoly Klypin and Octavio Valenzuela for
stimulating discussions, and for comments on a draft of this paper.
They openly shared their results ahead of publication and provided the
file of initial coordinates that we have used in these tests.  We also
thank an anonymous referee for a helpful report.  This work was
supported by grants from NASA (NAG 5-10110) and from NSF (AST-0098282)
to JAS.  VPD is supported by a Brooks Prize Fellowship in Theoretical
Physics at the University of Washington.

\appendix
\section*{Bar measurements}
We need to estimate the length of the bar from the simulations.  As
emphasized in previous studies, this is a non-trivial task because the
bar has no sharp edges, and the bi-symmetric density variations in the
simulation often include spirals and, less often, mildly distorted
rings.  Our procedure is motivated in part by that followed by
observers (\eg\ Aguerri \etal\ 2003), who have a smooth light
distribution instead of the noisy set of particles in our problem, but
who have only a single snapshot of an inclined disk.

Since we use a polar grid for the disk component, it is a simple
matter to save the amplitude and phase of the $m=2$ component of the
projected surface density on each radial ring at frequent intervals.
We find the pattern speed of the bar, $\Omega_p$, which we require to
be the same at all radii, from a fit to these coefficients in the
inner grid over a short time interval around the time at which we
desire a measurement.  We determine the corotation radius, $R_c$ from
the azimuthally averaged rotation curve at the mid-point in time of
the selected data.  We have not tried to determine the Lagrange point,
as did DS00, since the difference from the averaged corotation radius
is generally small.

Each fit for $\Omega_p$ also yields the radial profile of the relative
amplitude and phase of the $m=2$ density variations, which are
equivalent to aligning the data from each moment along a common axis
dictated by the fitted $\Omega_p$.  Averaging in this way both
diminishes shot noise from the particles and weakens the significance
of any other bi-symmetric feature, such as a spiral pattern, that may
rotate at a different pattern speed.  We use the ratio of the total
amplitude of the $m=2$, relative to the $m=0$ terms, from this
time-averaged density as a measure of the bar amplitude.

The radius at which the relative amplitude of the $m=2$ density is
half that of its peak value is generally quite stable over time, but
is clearly an underestimate of the bar length; unfortunately, we found
that the radius where the amplitude is only 10\% of the peak
fluctuates quite wildly because of imperfectly eliminated spiral arms
and the like.  We therefore persisted with this low estimate (a), the
radius of the half-peak amplitude, and sought a second high estimate
(b) which is the radius at which the phase shifts by $20^\circ$ from
its mass-weighted average value near the peak.  As this latter is
generally quite clearly an overestimate of the bar length, we define
the bar length, $a_B$, to be the average of estimates (a) and (b).

The dimensionless ratio ${\cal R} = R_c/a_B$, which can be compared
with observed values (\S\ref{intro}), is generally greater than unity.
Unfortunately, estimate (b) of $a_B$ suffers from large upward
fluctuations at times, which are only half eliminated by averaging,
and lead to corresponding downward fluctuations in $\cal R$, which are
readily recognized because we monitor the value continuously -- an
example is visible in Fig.~\ref{Rplot}.  Our quoted values of ${\cal
R}$ generally ignore large, short-lived downward fluctuations.  We
have found that bars in cool, massive disks generally form with ${\cal
R} \simeq 1$, as found in previous work, and that ${\cal R}$
fluctuates around unity in an extreme maximum disk model in which the
bar was hardly braked at all, giving us confidence in our estimation
method.


\begin{references}{}

\bigskip
\reference{}
Aguerri, J. A. L., Debattista, V. P. \& Corsini, E. M. 2003, \mnras, {\bf 338}, 465

\reference{}
Athanassoula, E. 1992, \mnras, {\bf 259}, 345

\reference{}
Athanassoula, E. 2002, \apjl, {\bf 569}, L83

\reference{}
Athanassoula, E. 2003, \mnras, {\bf 341}, 1179

\reference{}
Athanassoula, E. \& Sellwood, J. A. 1986, \mnras, {\bf 221}, 213

\reference{}
Binney, J. \& Tremaine, S. 1987, {\it Galactic Dynamics\/} (Princeton: Princeton University Press)

\reference{}
Buta, R. \& Combes, F. 1996, \fcp, {\bf 17}, 95

\reference{}
Debattista, V. P. \& Sellwood, J. A. 1998, \apjl, {\bf 493}, L5

\reference{}
Debattista, V. P. \& Sellwood, J. A. 2000, \apj, {\bf 543}, 704

\reference{}
Debattista, V. P. \& Williams, T. B. 2004, {\bf 605}, 714

\reference{}
Hernquist, L. 1990, \apj, {\bf 356}, 359

\reference{}
Hernquist, L. \& Weinberg, M. D. 1992, \apj, {\bf 400}, 80

\reference{}
Holley-Bockelmann, K. \& Weinberg, M. 2005, DDA abstract 36.0512

\reference{}
Kalnajs, A. J. 1977, \apj, {\bf 212}, 637

\reference{}
Klypin, A., Kravtsov, A. V., Valenzuela, O. \& Prada, F. 1999, \apj, {\bf 522}, 82

\reference{}
Kravtsov, A. V., Klypin, A. \& Khokhlov, A. M. 1997, \apjs, {\bf 111}, 73

\reference{}
Lin, D. N. C. \& Tremaine, S. 1983, \apj, {\bf 264}, 364

\reference{}
Little, B. \& Carlberg, R. G. 1991, \mnras, {\bf 250}, 161

\reference{}
Moore, B., Governato, F., Quinn, T., Stadel, J. \& Lake, G. 1998, \apjl, {\bf 499}, L5

\reference{}
Navarro, J. F., Frenk, C. S. \& White, S. D. M. 1997, \apj, {\bf 490}, 493

\reference{}
O'Neill, J. K. \& Dubinski, J. 2003, \mnras, {\bf 346}, 251

\reference{}
Rautiainen, P., Salo, H. \& Laurikainen, E. 2005, \apjl, {\bf 631}, L129

\reference{}
Sakamoto, K., Okamura, S. K., Ishizuki, S. \& Scoville, N. Z. 1999, \apj, {\bf 525}, 691

\reference{}
Sellwood, J. A. 1981, \aap, {\bf 99}, 362

\reference{}
Sellwood, J. A. 2003, \apj, {\bf 587}, 638

\reference{}
Sellwood, J. A. 2005, \apj, to appear (astro-ph/0407533) (Paper I)

\reference{}
Sellwood, J. A. \& Debattista, V. P. 2006, in preparation (Paper III)

\reference{}
Sellwood, J. A. \& Merritt, D. 1994, \apj, {\bf 425}, 530

\reference{}
Toomre, A. 1981, in {\it The Structure and Evolution of Normal Galaxies}, ed.\ S. M. Fall \& D. Lynden-Bell (Cambridge: Cambridge University Press), p.~111

\reference{}
Tremaine, S. \& Weinberg, M. D. 1984, \mnras, {\bf 209}, 729

\reference{}
Valenzuela, O. \& Klypin, A. 2003, \mnras, {\bf 345}, 406 (VK03)

\reference{}
Weinberg, M. D. 1985, \mnras, {\bf 213}, 451

\reference{}
Weinberg, M. D. 2004, \mnras, submitted (astro-ph/0404169)

\reference{}
Weinberg, M. D. \& Katz, N. 2002, \apj, {\bf 580}, 627

\reference{}
Weinberg, M. D. \& Katz, N. 2005, \mnras, submitted (astro-ph/0508166)

\end{references}
\end{document}